\documentclass[%
 amsmath,amssymb,
 aps,
prb,
twocolumn,
]{revtex4-2}

\usepackage{graphicx}
\usepackage{dcolumn}
\usepackage{bm}
\usepackage{multirow}
\usepackage[utf8]{inputenc}
\usepackage[T1]{fontenc}

\begin{document}
\title{Three-qubit entangling gates with simultaneous exchange controls in spin qubit systems}

\author{Miguel G. Rodriguez}
 \affiliation{Department of Physics, University of Texas at El Paso, El Paso, Texas 79968, USA }%
\author{Yun-Pil Shim}%
 \email{yshim@utep.edu}
 \affiliation{Department of Physics, University of Texas at El Paso, El Paso, Texas 79968, USA }%

\date{\today}

\begin{abstract}
Pairwise exchange couplings have long been the standard mechanism for entangling spin qubits in semiconductor systems. However, implementing quantum circuits based on pairwise exchange gates often requires a lengthy sequence of elementary gate operations. In this work, we present an alternative approach: multi-qubit entangling gate operations that simultaneously drive the exchange couplings between multiple pairs of spin qubits. We explore three spin qubit systems in linear or triangular configurations. We derive analytical expressions for these multi-exchange entangling operations and demonstrate how to use the resulting three-qubit gates to construct quantum circuits capable of generating standard entangled states such as GHZ and W states, and the Toffoli gate, by optimizing control parameters. Our results show that this multi-qubit strategy significantly reduces the number of required operations, offering a pathway to more efficient, shallower, and more coherent circuits for spin-qubit processors.
\end{abstract}

\maketitle

\section{Introduction}\label{sec:Introduction}
Semiconductor spin qubits have been a physical system that people have sought for a scalable qubit platform \cite{awschalom_science2013,kloeffel_annurevcmp2013,burkard_rmp2023}. Recent progress has led to semiconductor quantum dot devices with fidelities for gate operations comparable to those of other leading qubit platforms \cite{noiri_nat2022,xue_nat2022,mills_sciadv2022} and with an increasing number of spin qubits \cite{philips_nature2022,weinstein_nature2023,neyens_nat2024}. There are different types of spins in semiconductor systems that can be used as qubits, such as electron \cite{loss_divincenzo_pra1998}, nuclear \cite{kane_nature1998}, and hole \cite{scappucci_nrevmat2021,Fang_2023} spins. For all these spin qubits, the most common mechanism for two-qubit entangling gates between spin qubits is the exchange interaction, which can be easily controlled by electrical methods. Encoded qubits \cite{divincenzo_bacon_nature2000,RX_theory_Taylor_prl2013,RX_exp_Medford_prl2013,shim_tahan_prb2016,andrews_nnano2019,weinstein_nature2023,acuna_prappl2024} with multiple spins further take advantage of the fast and efficient exchange operations. 

However, the exchange interaction is very short-ranged. Implementing quantum circuits using short-ranged pairwise exchange couplings requires a long sequence of operations \cite{fong_wandzura_qic2011}, especially as the system size increases, limiting the device's applicability. For the further development of large-scale quantum computing devices based on spin qubits, a better approach is needed to generate entangling circuits for larger systems. Protocols for generating multi-qubit entangling gates in spin qubit systems have been proposed by simultaneously applying resonant driving and exchange interactions \cite{gullans_petta_prb2019} or by utilizing anisotropic chiral interactions \cite{nguyen_stefano_prxq2025}. 

In this work, we investigate the possibility of simple multi-qubit gates with simultaneous pairwise exchange interactions between multiple pairs of spin qubits. Simultaneous control of exchange couplings in spin qubit systems has gained increasing interest recently in the context of exchange-only qubits \cite{shim_tahan_prb2016,acuna_prappl2024,shim_apq2025,heinz_prxq2025,broz_arxiv2025}. Here, we consider conventional spin qubits and propose a simple protocol for generating multiqubit entangling gates by applying exchange interactions. Although any entangling two-qubit gate, along with general single-qubit gates, suffices to form a universal gate set \cite{divincenzo_pra95,barenco1995}, we show that the three-qubit entangling gate resulting from the simultaneous exchange control can significantly reduce the circuit depth and number of operations.  

We consider three spin qubits coupled by exchange interactions. Multiple exchange interactions are turned on and off simultaneously by controlling the voltages of a square pulse shape with different exchange strengths. An exact expression for the three-qubit entangling gate will be derived as a function of the control parameters, the qubit frequency, and the exchange couplings. We demonstrate the efficiency of the three-qubit entangling gate by composing the quantum circuits using the three-qubit gate and single-qubit gates, to generate some standard three-qubit entangled states and the standard Toffoli gate.

\section{Three-qubit entangling gate}\label{sec:U3_gate}

\subsection{System Hamiltonian and Time-Evolution Operator}
\begin{figure}
  \includegraphics[width=\linewidth]{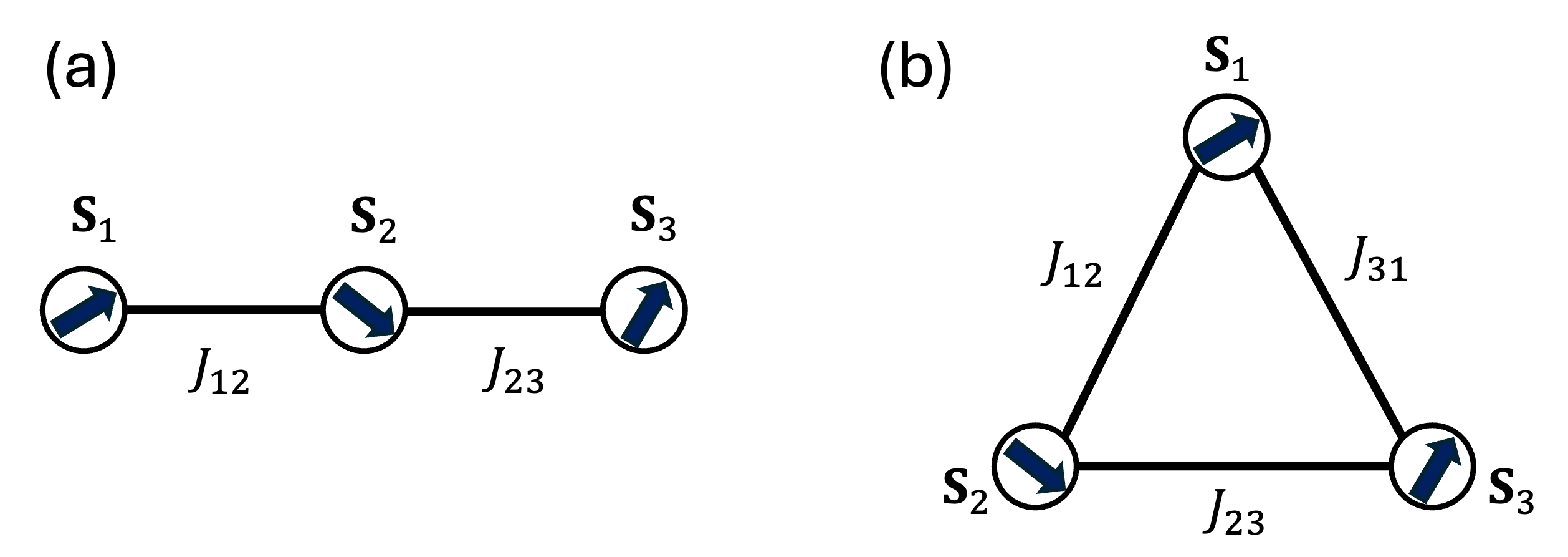}\\
  \caption{Two geometries of three spin qubit systems, (a) linear geometry and (b) triangular geometry. Each neighboring pair of spins are coupled by an exchange interaction.}
  \label{fig1:Geometry}
\end{figure}

We consider a system of three spin qubits with exchange couplings in either linear or triangular geometry (see Fig. \ref{fig1:Geometry}). The corresponding Hamiltonian is
\begin{equation}\label{eq:H}
\hat{H}= 
\varepsilon_1 S_1^z + \varepsilon_2 S_2^z + \varepsilon_3 S_3^z
+ J_{12} \mathbf{S}_1 \cdot \mathbf{S}_2
+ J_{23} \mathbf{S}_2 \cdot \mathbf{S}_3
+ J_{31} \mathbf{S}_3 \cdot \mathbf{S}_1 
\end{equation}
where $\varepsilon_i$ is the energy of the $i$-th quantum dot level and $J_{ij}$ is the exchange interaction between spins $\mathbf{S}_i$ and $\mathbf{S}_j$.
For the linear geometry, $J_{31}=0$. 

The general time evolution of the system with all the exchange interactions turned on leads to a rather complicated analytical expression. We are going to consider special cases where all QD energy levels are equal, $\varepsilon_1=\varepsilon_2=\varepsilon_3=\varepsilon$. In this case, the Hamiltonian commutes with the total spin operator $\mathbf{S}_{\mathrm{tot}}=\sum_i \mathbf{S}_i$, and we can use the total spin basis, which simplifies the Hamiltonian into a block-diagonal matrix with the largest block size of $2 \times 2$. The total spin basis $|S_{\mathrm{tot}}, S_{\mathrm{tot}}^z \rangle $ we used in this work is
\begin{eqnarray*}
|1\rangle &=& |S_{\mathrm{tot}}=\frac{3}{2},S_{\mathrm{tot}}^z=\frac{3}{2} \rangle = |\uparrow \uparrow \uparrow \rangle \\
|2\rangle &=& |\frac{3}{2},\frac{1}{2}\rangle = \frac{1}{\sqrt{3}} \left(|\uparrow \uparrow \downarrow \rangle + |\uparrow \downarrow \uparrow \rangle + |\downarrow \uparrow \uparrow \rangle \right)  \\
|3\rangle &=& |\frac{3}{2},-\frac{1}{2}\rangle = \frac{1}{\sqrt{3}} \left(|\uparrow \downarrow \downarrow \rangle + |\downarrow \uparrow \downarrow \rangle + |\downarrow \downarrow \uparrow \rangle \right) \\
|4\rangle &=&  |\frac{3}{2},-\frac{3}{2}\rangle = |\downarrow \downarrow \downarrow \rangle \\
|5\rangle &=& |\frac{1}{2}, \frac{1}{2}, \lambda=1 \rangle = \sqrt{\frac{2}{3}} |\uparrow \uparrow \downarrow \rangle - \frac{1}{\sqrt{6}} |\uparrow \downarrow \uparrow \rangle - \frac{1}{\sqrt{6}} |\downarrow \uparrow \uparrow \rangle \\
|6\rangle &=& |\frac{1}{2}, \frac{1}{2}, \lambda=2 \rangle = \frac{1}{\sqrt{2}} |\uparrow \downarrow \uparrow \rangle - \frac{1}{\sqrt{2}} |\downarrow \uparrow \uparrow \rangle \\
|7\rangle &=& |\frac{1}{2}, -\frac{1}{2}, \lambda=1 \rangle = \frac{1}{\sqrt{6}} |\uparrow \downarrow \downarrow \rangle + \frac{1}{\sqrt{6}} |\downarrow \uparrow \uparrow \rangle - \sqrt{\frac{2}{3}} |\downarrow \downarrow \uparrow \rangle \\
|8\rangle &=& |\frac{1}{2}, \frac{1}{2}, \lambda=2 \rangle = \frac{1}{\sqrt{2}} |\uparrow \downarrow \downarrow \rangle - \frac{1}{\sqrt{2}} |\downarrow \uparrow \downarrow \rangle
\end{eqnarray*}
where $\lambda = 1, 2$ represents the degenracy in $S_{\mathrm{tot}}=\frac{1}{2}$ states corresponding to $S_{12} = 1, 0$ for $\mathbf{S}_{12}=\mathbf{S}_1 + \mathbf{S}_2$, respectively.

The Hamiltonian matrix in the total spin basis is block diagonal with $1\times 1$ and $2\times 2$ blocks, with doubly degenerate $S_{\mathrm{tot}}=1/2$ states. The Hamiltonian matrix is
\begin{equation}\label{eq:Hmatrix}
\mathbb{H} = 
\left(
\begin{array}{cccccccc}
 a_1 & 0 & 0 & 0 & 0 & 0 & 0 & 0 \\
 0 & a_2 & 0 & 0 & 0 & 0 & 0 & 0 \\
 0 & 0 & a_3 & 0 & 0 & 0 & 0 & 0 \\
 0 & 0 & 0 & a_4 & 0 & 0 & 0 & 0 \\
 0 & 0 & 0 & 0 & b & c & 0 & 0 \\
 0 & 0 & 0 & 0 & c & d & 0 & 0 \\
 0 & 0 & 0 & 0 & 0 & 0 & b' & c \\
 0 & 0 & 0 & 0 & 0 & 0 & c & d'
\end{array}
\right) ~,
\end{equation}
where
\begin{eqnarray}
a_1 &=& \frac{3\varepsilon}{2} + \frac{1}{4}\left( J_{12} + J_{23} + J_{31} \right) \\
a_2 &=&  \frac{\varepsilon}{2} + \frac{1}{4}\left( J_{12} + J_{23} + J_{31} \right) \\
a_3 &=&  - \frac{\varepsilon}{2}\ + \frac{1}{4}\left( J_{12} + J_{23} + J_{31} \right) \\
a_4 &=&  - \frac{3\varepsilon}{2}\ + \frac{1}{4}\left( J_{12} + J_{23} + J_{31} \right) \\
b &=&  \frac{\varepsilon}{2} + \frac{1}{4}J_{12} - \frac{1}{2}J_{23} - \frac{1}{2}J_{31} \\
d &=&  \frac{\varepsilon}{2} - \frac{3}{4}J_{12} \\
b' &=&  - \frac{\varepsilon}{2} + \frac{1}{4}J_{12} - \frac{1}{2}J_{23} - \frac{1}{2}J_{31} \\
d' &=& - \frac{\varepsilon}{2} - \frac{3}{4}J_{12}  \\
c &=&  \frac{\sqrt{3}}{4}J_{23} - \frac{\sqrt{3}}{4}J_{31} ~.
\end{eqnarray}
To find the exact expression for the time evolution operator, we first find the eigenstates of the Hamiltonian. We can diagonalize the $2\times 2$ blocks to find the eigenstates and eigenvalues of the Hamiltonian.  In the basis of the eigenstates, the time-evolution operator is a trivial diagonal matrix with its diagonal elements $\exp\left( -i E_k t / \hbar\right)$ for each eigenvalue, $E_k$. Then we do the unitary transformation to the original computational basis to obtain the unitary time-evolution operator in the computational basis,
\begin{equation}\label{eq:U3}
   \mathbb{U}_{3} = \left(
\begin{array}{cccccccc}
 A & 0 & 0 & 0 & 0 & 0 & 0 & 0 \\
 0 & B & C & 0 & E & 0 & 0 & 0 \\
 0 & C & D & 0 & F & 0 & 0 & 0 \\
 0 & 0 & 0 & G' & 0 & F' & E' & 0 \\
 0 & E & F & 0 & G & 0 & 0 & 0 \\
 0 & 0 & 0 & F' & 0 & D' & C' & 0 \\
 0 & 0 & 0 & E' & 0 & C' & B' & 0 \\
 0 & 0 & 0 & 0 & 0 & 0 & 0 & A' 
\end{array}
\right)~.
\end{equation}
Explicit expressions for the matrix elements of the time-evolution operator are given in Appendix \ref{Sec:Append_U3}. This unitary operation represents the three-qubit entangling gate, $\mathbb{U}_{3}$.

\subsection{Construction of quantum circuits}
\begin{figure}
  \includegraphics[width=0.8\linewidth]{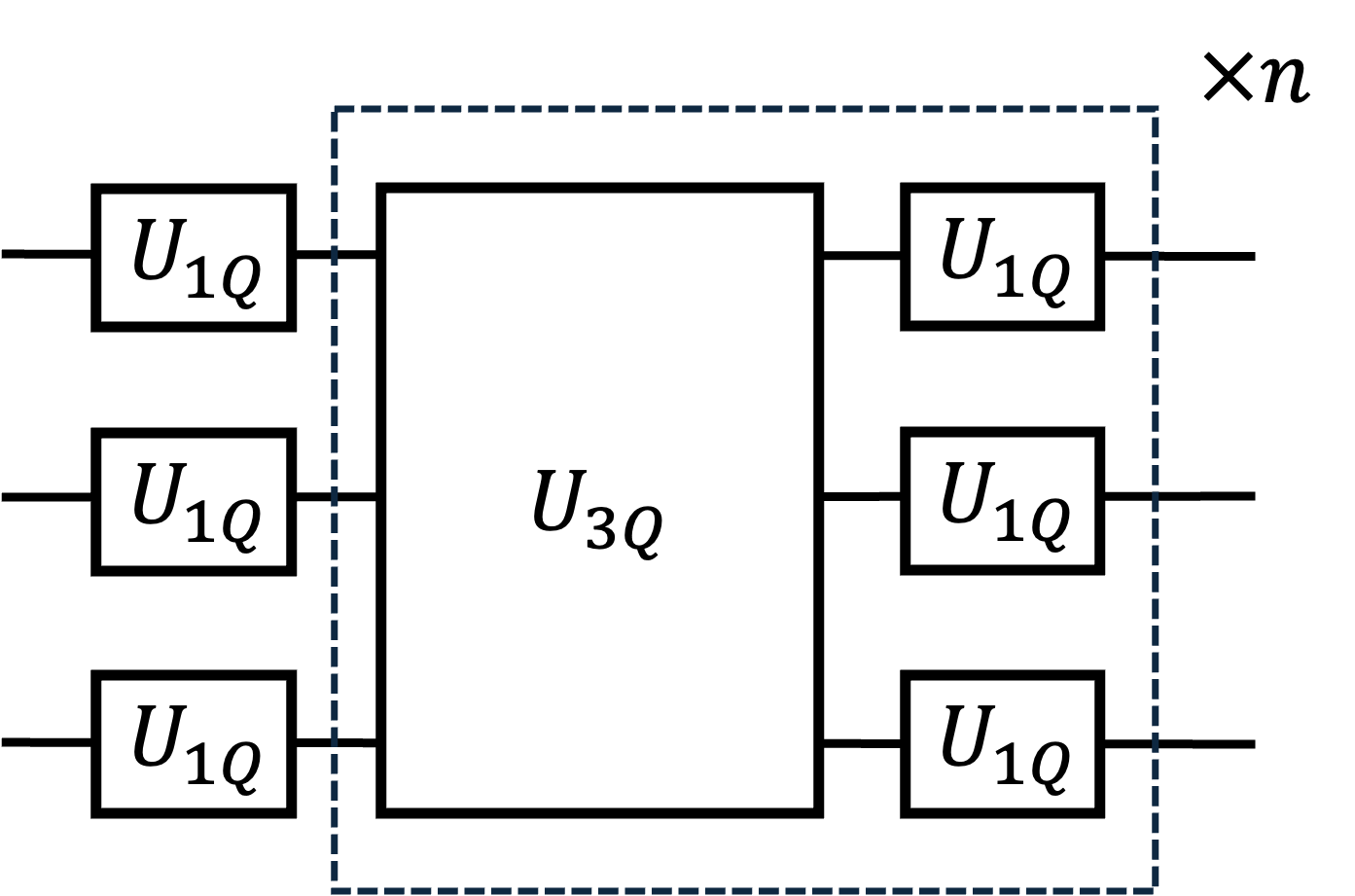}\\
  \caption{A quantum circuit utilizing the three-qubit entangling gate. The three-qubit gate is sandwiched by layers of single-qubit gates, and it is repeated as many times as necessary. All single-qubit gates and three-qubit gates will be individually optimized for the circuit to perform a given task.}
  \label{fig2:Multiqubit_gate_circuit}
\end{figure}

A common approach to composing a quantum circuit for a given task is to find a sequence consisting of single- and two-qubit gates, since they form a universal set of quantum gates \cite{divincenzo_pra95,barenco1995}. The three-qubit gate obtained in the previous section adds a new possibility of implementing quantum circuits. To investigate the usefulness and efficiency of the tree-qubit gate, we compose quantum circuits by using the three-qubit gates sandwiched by single-qubit gates, repeating as many times as necessary to implement a quantum circuit that performs a given task. Figure \ref{fig2:Multiqubit_gate_circuit} shows the general structure of the quantum circuit utilizing the three-qubit gates. 

The three-qubit gate, $\mathbb{U}_{3}$, can be tuned by tuning the control parameters, $\left\{\varepsilon, J_{12}, J_{23}, J_{31}, t \right\}$. We assume square pulse shape control of all the parameters during a turn-on time $t$. If multiple application of $\mathbb{U}_{3}$ is necessary, each $\mathbb{U}_{3}$ can be independently tuned. Single-qubit gates can be physically  implemented by electron spin resonance (ESR) \cite{koppens_prl2008,pla_nat2012,veldhorst_nat2015} or electric dipole spin resonance (EDSR) \cite{tokura_prl2006,pioro_ladriere_nphys2008,yoneda_nnanoh2018}. Here, we adopt three-step Euler rotations to represent each single-qubit gate, with the three Euler angles being the control parameters to be optimized, 
\begin{equation}
\hat{U}_{1Q}\left(\alpha,\beta,\gamma\right) = \hat{R}_z\left(\alpha\right) \hat{R}_y\left(\beta\right) \hat{R}_z\left(\gamma\right) ~,   
\end{equation}
where $\hat{R}_z$ and $\hat{R}_y$ are rotation operators around $z$ and $y$ direction, respectively. 
  
To find the correct single-qubit gates and the $\mathbb{U}_{3}$'s, we numerically optimize the circuit for a given task to find the optimal parameters. Each single-qubit layer has 9 parameters to optimize, and each $\mathbb{U}_{3}$ has four parameters, $\left\{ \tilde{\varepsilon}, \tilde{J}_{12}, \tilde{J}_{23}, \tilde{J}_{31} \right\}$, where $\tilde{\varepsilon}=\varepsilon t /\hbar$ and $\tilde{J}_{ij} = J_{ij} t / \hbar$, since we are using square pulses. We also need a global phase, which is mathematically necessary to find the optimal parameters that match the target circuit. The resulting circuit will therefore be equivalent to the target circuit up to a physically unimportant global phase. In total, if there are $n$ operations of $\mathbb{U}_{3}$, there are $13n + 10$ parameters to optimize for triangular geometry, and $12n + 10$ parameters for linear geometry since $\tilde{J}_{31}=0$. We used the BlackBoxOptim package \cite{BBO} to perform the global optimization.

\section{Applications}\label{sec:Applications}

\subsection{Generation of entangled states}

\begin{figure}
  \includegraphics[width=\linewidth]{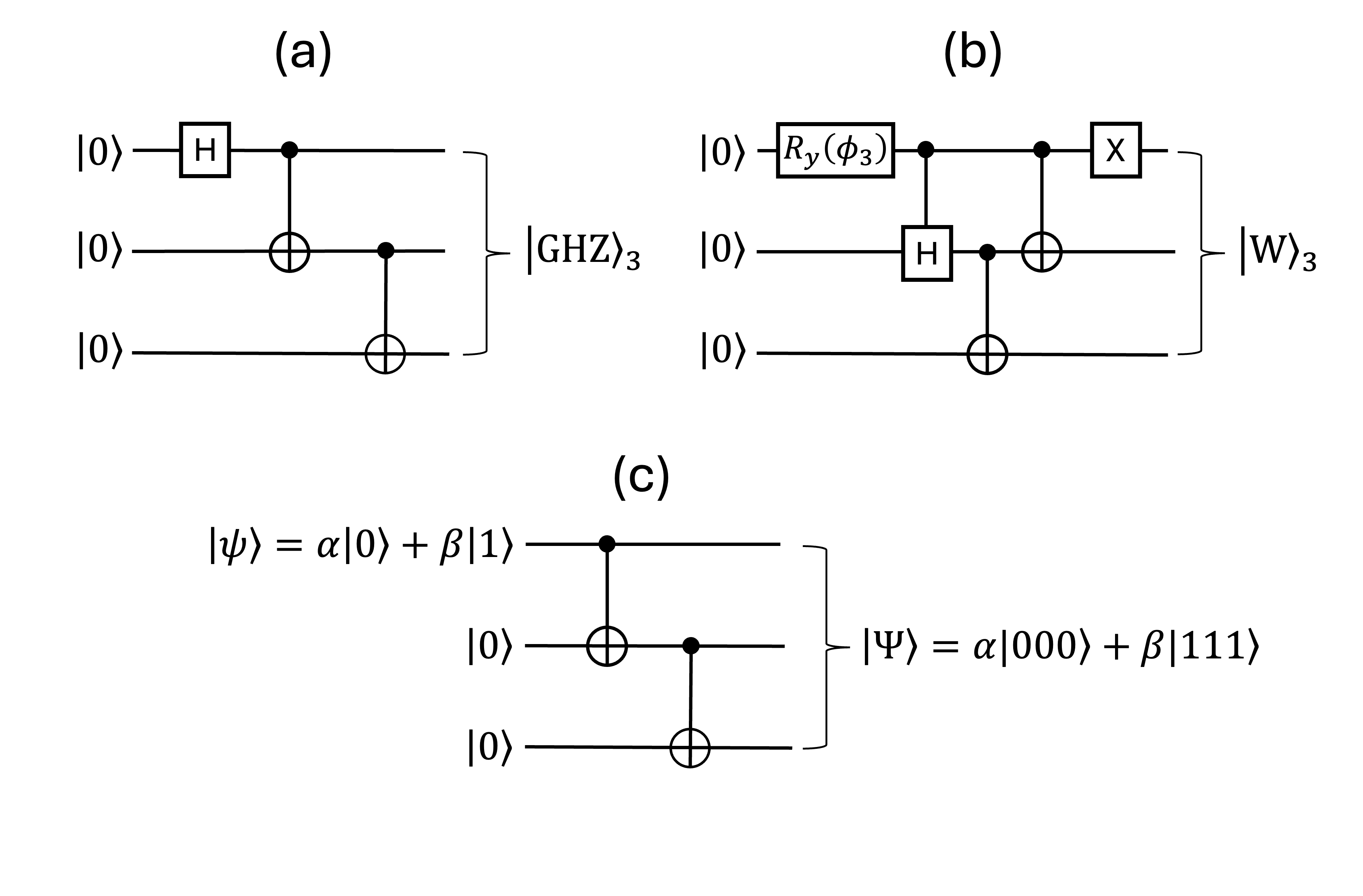}\\
  \caption{Standard quantum circuits for generating entangled three-qubit states. (a) A circuit for the three-qubit GHZ state, $|GHZ\rangle_3 = \left( |000\rangle + |111\rangle \right) / \sqrt{2}$. (b) A circuit for the three-qubit W state, $|W\rangle_3 = \left( |001\rangle + |010\rangle + |100\rangle \right) / \sqrt{3}$. $\phi_3=2\cos^{-1}({1/\sqrt{3}})$. (c) A circuit that encodes a three-qubit bit-flip error correction code.}
  \label{fig3:Standard_circuits}
\end{figure}

To demonstrate the efficiency of the three-qubit $\mathbb{U}_{3}$ gate, we tried to generate some standard entangled three-qubit states: GHZ \cite{GHZ1989_arxiv2007,GHZ1990} and W \cite{Wstate_PRA2000} states, using a standard approach and also using the $\mathbb{U}_{3}$ gate. Standard quantum circuits for these tasks involve single-qubit rotations and two-qubit CNOT gate operations [see Figs. \ref{fig3:Standard_circuits}(a) and (b)]. The CNOT gate can be implemented by various methods depending on the physical setup of the qubits. For simplicity, we consider a physical implementation of the CNOT gate with two $\sqrt{SWAP}$ operations and single-qubit gates \cite{loss_divincenzo_pra1998}. $\sqrt{SWAP}$ can be naturally generated by the exchange interactions between two spin qubits. We count the number of single-qubit steps and exchange operation steps for each quantum circuit and compare them with the necessary number of operation steps when we use the three-qubit gate as in Fig. \ref{fig2:Multiqubit_gate_circuit}.

\begin{table}
\begin{ruledtabular}
\begin{tabular}{c|cc|cc}
\multirow{2}{*}{Target state} & \multicolumn{2}{c|}{Standard circuit} & \multicolumn{2}{c}{$\mathbb{U}_3$ circuit} \\
  & $N_{\mathrm{exch}}$ & $N_{1Q}$ & $N_{\mathrm{exch}}$ & $N_{1Q}$ \\
\hline
\hline
GHZ & 4 & 5 & 1 & 2 \\
\hline
W & 4 & 5 & 1 & 2 \\ 
\hline
Bit-flip ECC & 4 & 5 & 3(L), 2(T) & 4(L), 3(T) 
\end{tabular}
\end{ruledtabular}
\caption{\label{tab:states}Two-qubit vs three-qubit gates for the generation of entangled three-qubit states. $N_{\mathrm{exch}}$ is the number of exchange steps, and $N_{1Q}$ is the number of signle-qubit steps in the circuits. For the bit-flip error-correction code, L denotes the linear geometry, and T denotes the triangular geometry.} 
\end{table}

For the GHZ state, the standard circuit [Fig. \ref{fig3:Standard_circuits}(a)] requires 4 exchange operations and 5 single-qubit steps. In contrast, the circuit in Fig. \ref{fig2:Multiqubit_gate_circuit} requires only a single operation of the three-qubit gate $\mathbb{U}_3$ ({\it i.e.}, $n=1$) along with two steps of the single-qubit gates to generate the GHZ state. For the W state, the standard circuit [Fig. \ref{fig3:Standard_circuits}(b)] requires 6 exchange operations and 7 single-qubit steps, taking into account that the controlled-Hadamard gate can also be implemented with two exchange operations ($\sqrt{SWAP}$). Using the three-qubit gate $\mathbb{U}_3$, it requires only a single operation of $\mathbb{U}_3$ and two single-qubit gate steps. Since the GHZ and W states are often used in many quantum algorithms, the multi-exchange three-qubit gate will lead to much more efficient implementation of the necessary quantum circuits. For generation of GHZ and W states, the number of required operations are equal in linear and triangular geometries. 

We also considered a quantum circuit that encodes a bit-flip error correcting code [Fig. \ref{fig3:Standard_circuits}(c)] which requires entangling three physical qubits. For any practical application of quantum computation, error correction would be essential to maintain the fault-tolerance of the quantum circuits. The bit-flip error-correction code would protect the encoded qubit information from any bit-flip errors on individual physical qubits. The standard circuit for encoding the bit-flip error-correcting code requires 4 exchange operations and 5 single-qubit steps. Using $\mathbb{U}_3$ in linear (triangular) geometry, we need three (two) operations of $\mathbb{U}_3$ and 4 (3) single-qubit steps, which is a modest improvement. We expect that the improvement would be more significant for encoding a full error-correction code.

\subsection{Generation of Toffoli gate}

\begin{figure}
  \includegraphics[width=\linewidth]{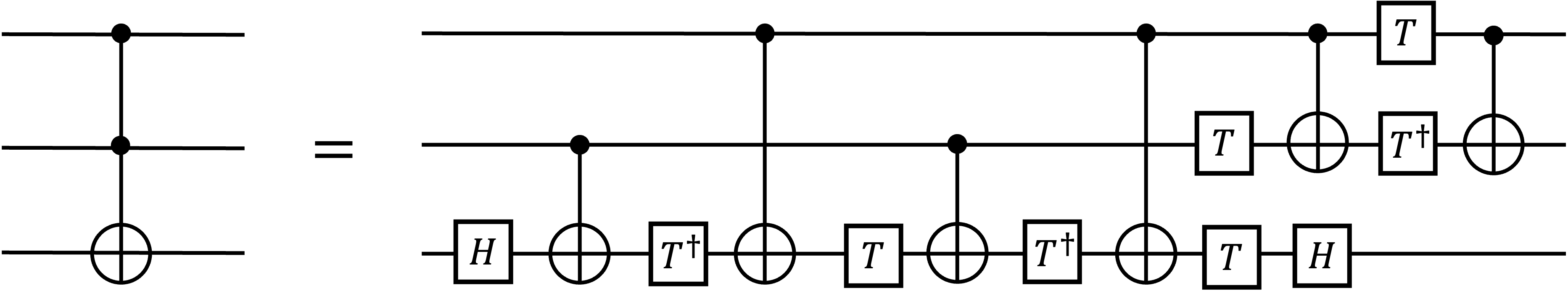}\\
  \caption{A quantum circuit that implements the Toffoli gate using single-qubit gates and two-qubit CNOT gates \cite{Nielsen_Chuang,Toffoli_circuit}.}
  \label{fig4:Toffoli}
\end{figure}

\begin{table}
\begin{ruledtabular}
\begin{tabular}{c|cc|cc}
\multirow{2}{*}{Geometry} & \multicolumn{2}{c|}{Standard circuit} & \multicolumn{2}{c}{$\mathbb{U}_3$ circuit} \\
  & $N_{\mathrm{exch}}$ & $N_{1Q}$ & $N_{\mathrm{exch}}$ & $N_{1Q}$ \\
\hline
\hline
Linear & 16 & 17 & 8 & 9 \\
\hline
Triangular & 12 & 13 & 7 & 8  
\end{tabular}
\end{ruledtabular}
\caption{\label{tab:Toffoli}Two-qubit vs three-qubit gates for generation of Toffoli gate. In linear geometry, CNOT gate bewteen two non-neighboring qubits requires additional SWAP operations. } 
\end{table}

In addition to generating entangled three-qubit states, the three-qubit gate $\mathbb{U}_3$ can be used to efficiently generate other multi-qubit entangling gates. As an example, we use the circuit in Fig. \ref{fig2:Multiqubit_gate_circuit} to generate the Toffoli gate \cite{Toffoli}. The Toffoli gate is important since it can form a universal set of quantum gates with single-qubit gates. Efficient composition of the Toffoli gate using single- and two-qubit gates has been an important task since the early days of quantum computing \cite{barenco_pra1995,sleator_prl1995}. Implementation of the Toffoli gate with only the two-qubit CNOT gate and single-qubit gates requires at least five two-qubit (or six CNOT) gates \cite{yu_pra2013} (Fig. \ref{fig4:Toffoli}), which would require a long sequence of physical operations to implement the necessary single- and two-qubit gates. In linear geometry, it requires 16 exchange operations (including the necessary SWAP operations between qubit 1 and qubit 2) and 17 single-qubit steps. In triangular geometry, the standard circuit requires 12 exchange operations and 13 single-qubit steps. 

$\mathbb{U}_3$ can significantly reduce the depth of the quantum circuit. In linear geometry, we need 8 operations of $\mathbb{U}_3$ and 9 single-qubit steps to implement the Toffoli gate. In triangular geometry, we can implement the Toffoli gate with only 7 operations of $\mathbb{U}_3$ and 8 single-qubit steps.

\section{Conclusion}
We have presented a promising way to design a three-qubit entangling gate in spin qubit systems based on the simultaneous control of the exchange interactions between spin qubits.
We demonstrated its effectiveness in generating some of the standard entangled states and entangling gates for three-qubit systems. In all the cases we considered here, the use of the three-qubit $\mathbb{U}_3$ gate significantly reduced the depth of the necessary quantum circuits. Although we considered simple examples with three-qubit systems, it is expected that utilizing the entangling power of generically multi-qubit gates would reduce the number of steps and, therefore, the strict requirements on the coherence of the qubits and the fidelities of the gate operations. 

Here we investigated only a particular way of using the multi-qubit gate, shown in Fig. \ref{fig2:Multiqubit_gate_circuit}. For more complicated quantum circuits, all available single-, two- three-qubit gates can be combined to efficiently generate the target quantum circuit, although it would require more complicated optimization to find the optimal composition and sequence.    

Simultaneous control of the pairwise exchange couplings can be extended to more than three-qubit systems to generate general multi-qubit entangling gates. We are working on extending this to a four-qubit system. This approach can also be applied to other physical qubit systems, such as superconducting qubits \cite{devoret_science2013,kjaergaard_arcmp2020}. Superconducting qubits have different types of interactions. For example, transmon qubits \cite{transmon} with capacitive couplings have anisotropic exchange interactions \cite{krantz_apr2019}. We expect that simultaneously turning on the pairwise qubit-qubit interactions would lead to multi-qubit entangling gates, which can potentially be used for efficient implementation of quantum circuits.

\begin{acknowledgments}
Research was sponsored by the Army Research Office (ARO) and was accomplished under Cooperative Agreement Number W911NF-24-2-0082. The views and conclusions contained in this document are those of the authors and should not be interpreted as representing the official policies, either expressed or implied, of the Army Research Office or the U.S. Government. The U.S. Government is authorized to reproduce and distribute reprints for Government purposes notwithstanding any copyright notation herein.
This work was also supported by the Air Force Office of Scientific Research (AFOSR) through grant No. FA9550-23-1-0477 and by the National Science Foundation (NSF) under Award Number 2316808. 
\end{acknowledgments}

\begin{widetext}

\appendix

\section{3-spin time-evolution operator}\label{Sec:Append_U3}
The time-evolution of the Hamiltonian in Eq. (\ref{eq:H}) in the computational basis can be exactly found. We can find the eigenvalues and eigenvectors of the Hamiltonian in the total spin basis since the Hamiltonian becomes block-diagonal with the largest block of size $2\times 2$. Then in the eigenbasis, the time evolution is a trivial diagonal matrix with its elements $\exp\left( -i E_k t / \hbar\right)$ for each eigenvalue, $E_k$. Then we do the unitary transformation to the original computational basis to obtain the unitary time-evolution operator in the computational basis, give by Eq. (\ref{eq:U3}):
\begin{equation}
   \mathbb{U}_{3} = \left(
\begin{array}{cccccccc}
 A & 0 & 0 & 0 & 0 & 0 & 0 & 0 \\
 0 & B & C & 0 & E & 0 & 0 & 0 \\
 0 & C & D & 0 & F & 0 & 0 & 0 \\
 0 & 0 & 0 & G' & 0 & F' & E' & 0 \\
 0 & E & F & 0 & G & 0 & 0 & 0 \\
 0 & 0 & 0 & F' & 0 & D' & C' & 0 \\
 0 & 0 & 0 & E' & 0 & C' & B' & 0 \\
 0 & 0 & 0 & 0 & 0 & 0 & 0 & A' 
\end{array}
\right)~.
\end{equation}
The elements are
\begin{eqnarray*}
A &=& \gamma_1 \\
B &=& \frac{1}{3}\left[\gamma_2+\gamma_5+\gamma_6+\left(\gamma_5-\gamma_6 \right)\cos\theta  \right] \\
C &=& \frac{1}{6}\left[2\gamma_2-\gamma_5-\gamma_6 - \left(\gamma_5-\gamma_6 \right) \left(  \cos\theta - \sqrt{3}\sin\theta \right) \right] \\
D &=& \frac{1}{6}\left[2 \left(\gamma_2+\gamma_5+\gamma_6\right) - \left(\gamma_5-\gamma_6 \right) \left( \cos\theta + \sqrt{3}\sin\theta \right) \right] \\
E &=& \frac{1}{6}\left[2\gamma_2-\gamma_5-\gamma_6 - \left(\gamma_5-\gamma_6 \right) \left( \cos\theta + \sqrt{3}\sin\theta \right) \right] \\
F &=& \frac{1}{6}\left[2\gamma_2-\gamma_5-\gamma_6+2\left(\gamma_5-\gamma_6 \right)\cos\theta  \right] \\
G &=& \frac{1}{6}\left[2 \left(\gamma_2+\gamma_5+\gamma_6\right) - \left(\gamma_5-\gamma_6 \right) \left( \cos\theta - \sqrt{3}\sin\theta \right) \right] 
\end{eqnarray*}
\begin{eqnarray*}
A' &=& \gamma_4 \\
B' &=& \frac{1}{3}\left[\gamma_3+\gamma_7+\gamma_8+\left(\gamma_7-\gamma_8 \right)\cos\theta  \right] \\
C' &=& \frac{1}{6}\left[2\gamma_3-\gamma_7-\gamma_8 - \left(\gamma_7-\gamma_8 \right) \left(  \cos\theta - \sqrt{3}\sin\theta \right) \right] \\
D' &=& \frac{1}{6}\left[2 \left(\gamma_3+\gamma_7+\gamma_8\right) - \left(\gamma_7-\gamma_8 \right) \left( \cos\theta + \sqrt{3}\sin\theta \right) \right] \\
E' &=& \frac{1}{6}\left[2\gamma_3-\gamma_7-\gamma_8 - \left(\gamma_7-\gamma_8 \right) \left( \cos\theta + \sqrt{3}\sin\theta \right) \right] \\
F' &=& \frac{1}{6}\left[2\gamma_3-\gamma_7-\gamma_8+2\left(\gamma_7-\gamma_8 \right)\cos\theta  \right] \\
G' &=& \frac{1}{6}\left[2 \left(\gamma_3+\gamma_7+\gamma_8\right) - \left(\gamma_7-\gamma_8 \right) \left( \cos\theta - \sqrt{3}\sin\theta \right) \right] 
\end{eqnarray*}
where $\gamma_k=\exp(-iE_k t /\hbar)$ with the eigenvalues $E_k$'s of the Hamiltonian,
\begin{eqnarray*}
E_1 &=& \frac{3\varepsilon}{2} + \frac{1}{4}\left( J_{12} + J_{23} + J_{31} \right) \\
E_2 &=&  \frac{\varepsilon}{2} + \frac{1}{4}\left( J_{12} + J_{23} + J_{31} \right) \\
E_3 &=&  - \frac{\varepsilon}{2}\ + \frac{1}{4}\left( J_{12} + J_{23} + J_{31} \right) \\
E_4 &=&  - \frac{3\varepsilon}{2}\ + \frac{1}{4}\left( J_{12} + J_{23} + J_{31} \right) \\
E_5 &=& \frac{\varepsilon}{2} - \frac{1}{4}\left( J_{12} + J_{23} + J_{31} \right)  + \frac{\sqrt{K}}{2} \\
E_6 &=& \frac{\varepsilon}{2} - \frac{1}{4}\left( J_{12} + J_{23} + J_{31}  \right) - \frac{\sqrt{K}}{2} \\
E_7 &=& -\frac{\varepsilon}{2} - \frac{1}{4}\left( J_{12} + J_{23} + J_{31} \right)  + \frac{\sqrt{K}}{2} \\
E_8 &=& -\frac{\varepsilon}{2} - \frac{1}{4}\left( J_{12} + J_{23} + J_{31} \right)  - \frac{\sqrt{K}}{2} 
\end{eqnarray*}
where $K=J_{12}^2+J_{23}^2+J_{31}^2-J_{12}J_{23}-J_{23}J_{31}-J_{31}J_{12}$, and the angle $\theta$ is defined as
\begin{equation}
\cos\theta = \frac{2J_{12}-J_{23}-J_{31}}{2\sqrt{K}} 
\end{equation}
in the range $\theta \in [0,\pi]$.

%
%

\end{widetext}

%
%
%
%

\bibliography{ref}

\end{document}